\documentclass[12pt]{article}
\usepackage{graphicx}
\usepackage[utf8]{inputenc}
\usepackage[english
]{babel}
\def\baselinestretch{1.5}
\begin{document}
\begin{center}
\bf{TRIANGLE GEOMETRY OF THE QUBIT STATE 
IN THE PROBABILITY REPRESENTATION EXPRESSED 
IN TERMS OF THE TRIADA OF MALEVICH'S  SQUARES}\\
\end{center}
\bigskip

\begin{center} {\bf V. N. Chernega$^1$, O. V. Man'ko$^{1,2}$, V. I. Man'ko$^{1,3}$}
\end{center}

\medskip

\begin{center}
$^1$ - {\it Lebedev Physical Institute, Russian Academy of Sciences\\
Leninskii Prospect 53, Moscow 119991, Russia}\\
$^2$ - {\it Bauman Moscow State Technical University\\
The 2nd Baumanskaya Str. 5, Moscow 105005, Russia}\\
$^3$ - {\it Moscow Institute of Physics and Technology (State University)\\
Institutskii per. 9, Dolgoprudnyi, Moscow Region 141700, Russia}\\
Corresponding author e-mail: manko@sci.lebedev.ru
\end{center}














\section*{Abstract}
We map the density matrix of the qubit (spin-1/2) state associated
with the Bloch sphere and given in the tomographic probability
representation onto vertices of a triangle determining {\it Triada
of Malevich's squares}. The three triangle vertices are located on
three sides of another equilateral triangle with the sides equal to
$\sqrt 2$. We demonstrate that the triangle vertices are in
one-to-one correspondence with the points inside the Bloch sphere
and show that the uncertainty relation for the three probabilities
of spin projections $+1/2$ onto three orthogonal directions has the
bound determined by the triangle area introduced. This bound is
related to the sum of three {\it Malevich's square} areas where the
squares have sides coinciding with the sides of the triangle. We
express any evolution of the qubit state as the motion of the three
vertices of the triangle introduced and interpret the gates of qubit
states as the semigroup symmetry of the {\it Triada of Malevich's
squares}. In view of the dynamical semigroup of the qubit-state
evolution, we constructed nonlinear representation of the group
$U(2)$.

\medskip

\noindent{\bf Keywords:} qubit state, Triada of Malevich's
squares, tomographic probability representation, positive map.

\section{Introduction}
The quantum states are described either by the wave function (pure
states)~\cite{Schr26} or the density
matrix~\cite{Landau27,vonNeumann27}. The spin-1/2 observables are
determined by the Pauli matrices, and the theory of the spin states
was suggested in \cite{Pauli}.  The spin-1/2 states are described by
the density 2$\times$2 matrices $\rho$ such that $\rho^\dag=\rho$,
$\mbox{Tr}\rho=1$, and the eigenvalues of the density matrices are
nonnegative. The observables of the spin-1/2 systems (or qubit
systems) are associated with the Pauli matrices $\sigma_x$,
$\sigma_y$, and $\sigma_z$ corresponding to spin projections
$m=\pm1/2$ on three orthogonal directions along the axes $x$, $y$,
and $z$, respectively. The tomographic probability representation of
spin states, where the states are described by fair probability
distributions, was introduced in \cite{DodPLA,OlgaJETP} and studied
in
\cite{Weigert1,Wiegert2,Painini,Bregence,Filipov,CastanosJPA,Klimov,OlgaWigner}.

The standard geometrical picture of qubit states is associated with
the Bloch-sphere points. The points on the surface of the Bloch
sphere correspond to the pure qubit states. The points inside the
Bloch sphere correspond to the mixed qubit states. Recently
\cite{Patrizia}, in view of the tomographic probability
representation, the qubit states were associated with the points in
the sphere with radius $1/2$ and the center of the sphere coinciding
with the center of a cube with coordinates $x=1/2$, $y=1/2$, and
$z=1/2$, the cube side being equal to unity.

In \cite{Patrizia}, the explicit expression for the matrix elements
of the qubit density matrix $\rho$ was obtained in terms of three
probabilities $p_j$ $(1\geq p_j\geq 0$, $j=1,2,3)$ to have in this
state the spin projections $m=+1/2$ on three orthogonal directions
along the axes $x$, $y$, and $z$. Due to this tomographic
probability representation of the qubit density matrix $\rho$, the
inequality corresponding to quantum correlations of these spin
projections of the form
\begin{equation}\label{intq}
\sum_{j=1}^3(p_j-1/2)^2\leq 1/4
\end{equation}
was obtained. Also in \cite{Patrizia} it was proposed to check this
inequality (uncertainty relation) in the experiments with
superconducting qubits, where the qubit states are realized in the
devices based on Josephson
junctions~\cite{Shalibo,Martinis,Astafiev,Ustinov}.

The above-described geometrical representation of the qubit states
makes obvious the difference of the quantum spin-1/2 states
expressed in terms of the spin-projection probabilities and the
geometrical representation of the states of three classical coins
also described by three probabilities $p_j$ to have the coins in
positions ``up.'' The states of these three classical coins are
associated with all points in a cube with the side equal to unity,
since for classical coins there is no quantum constraint expressed
as an inequality for the three probabilities $p_1$, $p_2$, and
$p_3$. For quantum spin-1/2 states, the points outside of the sphere
but still inside of the cube are not realized.

The aim of this work is to propose another geometrical
interpretation of the qubit states as well as the states of three
classical coins. Namely, we suggest to associate the classical coin
states and the spin-1/2 states with three vertices of the triangle.
These three vertices are located on three different sides of another
triangle, which has equal sides of length $\sqrt 2$. Thus, we map
the points in a three-dimensional picture of the Bloch sphere and
the cube onto points on a plane. It provides the possibility to
associate the characteristics of the qubit states expressed in terms
of quadratic forms of the probabilities $p_j$ with the areas of
triangles and squares on the plane.

We introduce the idea of {\it Triada of Malevich's squares} for
identification of the spin-1/2 states in the triangle geometrical
picture of the qubit density matrix. The difference of the areas for
qubit states and for three classical coin states can characterize
the difference of classical and quantum correlations in the
elaborated geometric picture of the systems under discussion.

This paper is organized as follows.

In Sec. 2, we review the probability representation of spin states.
In Sec.~3, we consider the spin-1/2 density matrix using the
tomographic probability distributions. In Sec.~4, we introduce the
triangle geometric representation of qubit states. In Sec.~5, we
describe quantum correlations in the geometric representation. We
give our conclusions in Sec.~6.

\section{Spin Tomography}   
The states of spin-$j$ systems (states of qudits) are identified
with the Hermitian density operators $\hat\rho$. The $N$$\times $$N$
matrices $\rho$ of the operators in the $|m\rangle$ basis, i.e.,
$\rho_{m m'}=\langle m|\hat\rho| m'\rangle$, where $N=2j+1,$
$j=0,1/2,1,3/2,\ldots$, and $m, m'=-j,-j+1,\ldots,j-1,j$, are such
that $\rho^\dagger=\rho,$ $\mbox{Tr}\,\rho=1$, and $\rho\geq0$. The
eigenvalues of the density matrices are nonnegative. In
\cite{DodPLA,OlgaJETP}, the spin tomographic probability
distribution,
called the spin tomogram $w(m,{\bf n})\geq 0$, where $m$ is the spin
projection on the direction determined by the unit vector ${\bf
n}=(\sin\beta\cos\alpha,\,\sin\beta\sin\alpha,\,\cos\beta)$, was
introduced and suggested to be identified with the spin state. The
spin tomogram is defined in terms of the density matrix $\rho_{m
m'}$ as the diagonal matrix element of the density matrix in the
rotated reference frame $|m,u\rangle=u^{\dagger}|m\rangle$, where
the unitary matrix $u$ depends on the Euler angles
$\alpha$, $\beta$, $\gamma$,
\begin{equation}\label{eq.1}
w(m,{\bf n})=(u\rho u^{\dagger})_{m m}.
\end{equation}
The matrix $u$ is the unitary matrix of an irreducible
representation of the rotation group or the $SU(2)$ group. Since
there exists a one-to-one correspondence between the density matrix
$\rho_{m m'}$ and the probability distribution $w(m,{\bf
n})$~\cite{OlgaJETP}, the information on the spin system state
contained in the spin tomogram is identical to the information
contained in the density matrix $\rho_{m m'}$. The spin tomography
was studied in
\cite{Weigert1,Wiegert2,Painini,Filipov,CastanosJPA,OlgaMarmoPhysScripta}.
The relation of the spin tomography to the star-product quantization
schemes~\cite{Stratonovich} was discussed in
\cite{OlgaMarmoJPA,OlgaPatrizia,PatriziaLizzi,Ibort,150MarmoPScr}.
The geometric metric properties providing the description of the
distance between different qudit states were studied in
\cite{Patrizia}. For example, in \cite{Patrizia} the qubit density
matrix was presented in terms of the Pauli matrices and the three
probabilities $p_1$, $p_2$, and $p_3$ as follows:
\begin{equation}\label{eq.2}
\rho=\left[\sigma_0+\sum_{k=1}^3\left(2p_k-1\right)\sigma_k\right]/2,
\end{equation}
where $\sigma_0$ is the unity matrix and
\[\sigma_1=\left(\begin{array}{cc}
0&1\\
1&0
\end{array}\right),\qquad \sigma_2=\left(\begin{array}{cc}
0&-i\\
i&0
\end{array}\right), \qquad \sigma_3=\left(\begin{array}{cc}
1&0\\
0&-1
\end{array}\right), \qquad 1,2,3\equiv x,y,z.
\]
Probabilities $1\geq p_1,p_2,p_3\geq 0$ are the probabilities to
have in the state $\rho$ the spin projections $m=+1/2$ on the
directions $x$, $y$, and $z$, respectively. Formula~(\ref{eq.2}) is
connected with the Bloch-sphere representation of the qubit state.
The qubit density matrix in this representation reads
\begin{equation}\label{eq.3}
\rho=\left(\begin{array}{cc}
({1+z})/{2}&({x-i y})/{2}\\
({x+i y})/{2}&({1-z})/{2}
\end{array}\right).
\end{equation}
The nonnegativity condition of the density matrix provides the
inequality
\begin{equation}\label{eq.3a}
x^2+y^2+z^2\leq1.
\end{equation}
The parameters $x$, $y$, and $z$ are used to associate the qubit
states with the points either on the surface of the Bloch sphere
(pure states) or inside the Bloch sphere (mixed states). Thus, the
qubit states are known to have a geometrical interpretation in terms
of points associated with the Bloch sphere. Formula~(\ref{eq.2})
relates the qubit states with probabilities. It gives the
possibility to suggest another geometrical interpretation which we
present in the next section.

\section{Three Classical Coins Statistics and Triangle Geometry of Qubit States}
To elucidate the proposed triangle-geometry picture of qubit states,
we recall the statistical properties of three independent classical
coins. They are associated with three probability distributions
since we assume that the classical coins are not correlated. The
probability distribution for the first coin is given by nonnegative
numbers $p_1$ and $1-p_1=p_1'$. For the second coin, one has numbers
$p_2$ and $1-p_2=p_2'$. For the third coin one has numbers $p_3$ and
$1-p_3=p_3'$. The probabilities $p_k$ $(k=1,2,3)$ describe results
of the experiments, where the $k{\mbox{th}}$ coin looks ``up.'' The
numbers $p_k$ and $p_k'$ can be considered as components of the
$k{\mbox{th}}$ probability vector $~{\bf
p}_k=\left(\begin{array}{cc}
p_k\\
p_{k'}
\end{array}\right).$
Geometrically this vector can be presented on a plot; see Fig.~1.
\begin{figure}
  \includegraphics[width=150mm]{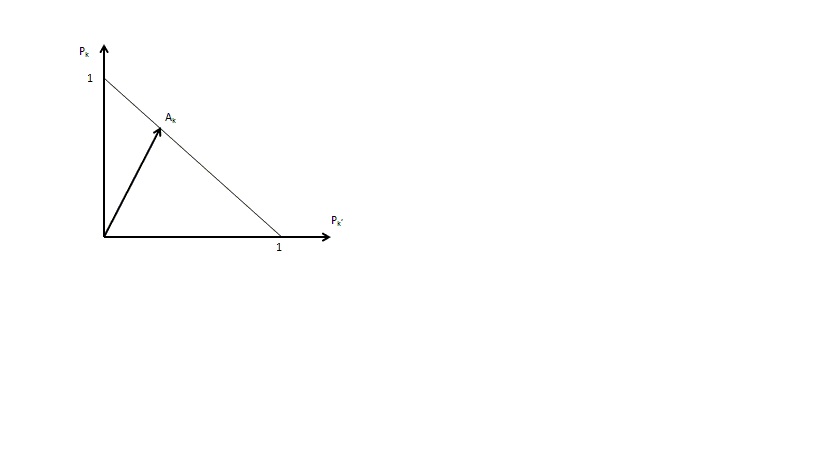}\\
  \caption{The probability vector ${\bf p}_k$ with the end at point $A_k$ on 
the simplex line determined by the equality $p_k+p_{k'}=1$.}
\end{figure}
In Fig.~1, the end of the vector ${\bf p}_k$ coincides with the
point $A_k$ on the line determined by the equation $p_k+p_{k'}=1$.
Due to the nonnegativity of numbers $p_k$ and $1-p_k$, the point
belongs to the simplex. The length of the simplex line is equal to
$\sqrt2$. Since we have three classical coins and three probability
vectors ${\bf p}_k$, the points with coordinates $p_1$, $p_2$, and
$p_3$ may be considered as the points either on the cube surface or
inside the cube in the three-dimensional space. The length of the
cube side is equal to unity.

There exists another possibility to provide the geometrical picture
of the three-coin probabilities. The three simplex lines can be
considered as the three sides of an equilateral triangle on the
plane of equal sides $\sqrt 2$; see Fig.~2.
\begin{figure}
 \includegraphics[width=150mm]{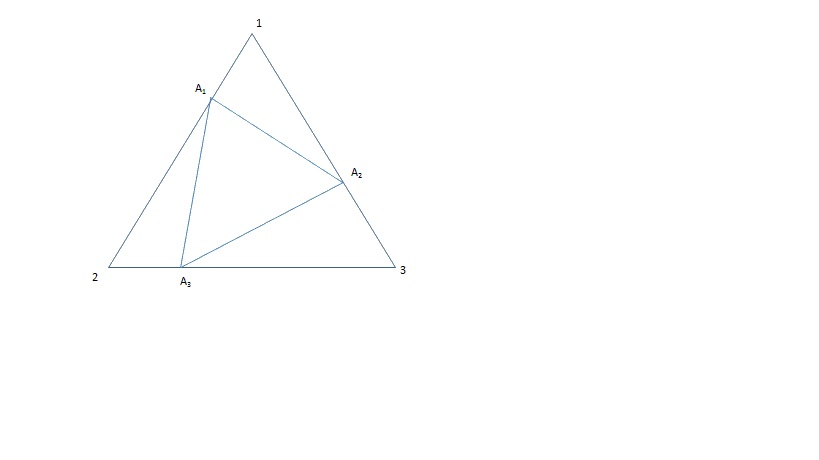}\\
  \caption{The equilateral triangle with vertices $1$, $2$, and $3$ and side 
length $\sqrt 2$ and vertices $A_1$,  $A_2$, and  $A_3$ determining the qubit state
}
\end{figure}
Thus, the points related to the cube are mapped onto the three
points $A_1$, $A_2$, and $A_3$ on the sides of the equilateral
triangle. One can connect these points by dashed lines and obtain a
triangle $A_1A_2A_3$ with vertices $A_1$, $A_2$, and $A_3$ located
on simplexes. This triangle can coincide with the equilateral
triangle. Also the sides of this triangle can have arbitrarily small
length and arbitrarily small area, if the points $A_1$ and $A_2$ or
$A_2$ and $A_3$ or $A_1$ and $A_3$ are very close each to the other.

\section{The Uncertainty Relation for Probabilities}    
For the qubit state determined by the density matrix $\rho$, the
explicit form of the density matrix expressed in terms of the
probabilities $p_k$ reads
\begin{equation}\label{eq.4}
\rho=\left(\begin{array}{cc}
p_3&p_1-ip_2-({1}/{2})+({i}/{2})\\
p_1+ip_2-({1}/{2})-({i}/{2})&1-p_3
\end{array}\right).
\end{equation}
In terms of the probabilities $p_k$. two eigenvalues of the matrix
$\rho$ are
$$\lambda_{1,2}=\frac{1}{2}\pm\left[\sum_{k=1}^3\left(
p_k-1/2\right)^2\right]^{1/2};
$$
they determine the Shannon entropy $H=-\lambda_1\ln\lambda_1
-\lambda_2\ln\lambda_2.$

For the density matrix, we describe the properties of the triangle
with vertices $A_1$, $A_2$, and $A_3$ located inside the equilateral
triangle with the side lengths equal to $\sqrt2$. We assume that the
vertices $A_k$ are closer to the $k{\mbox{th}}$ vertices of the
equilateral triangle, as shown in Fig.~2. The points $A_k$ have the
distance $d_k$ from the $k{\mbox{th}}$ vertices equal to
$d_k=p_k\sqrt2.$
Since the length of the side of the equilateral triangle is equal to
$\sqrt2$, one can calculate the lengths of the triangle sides $y_k$
of the triangle $A_1A_2A_3$; they are
\begin{equation}\label{eq.M2}
y_k=\left(2+2p_k^2-4p_k-2p_{k+1}+2p_{k+1}^2+2p_k p_{k+1}\right)^{1/2}.
\end{equation}
Now we construct three squares with sides $y_k$ associated with
triangle $A_1A_2A_3$ as shown in Fig.~3.
\begin{figure}
  \includegraphics[width=150mm]{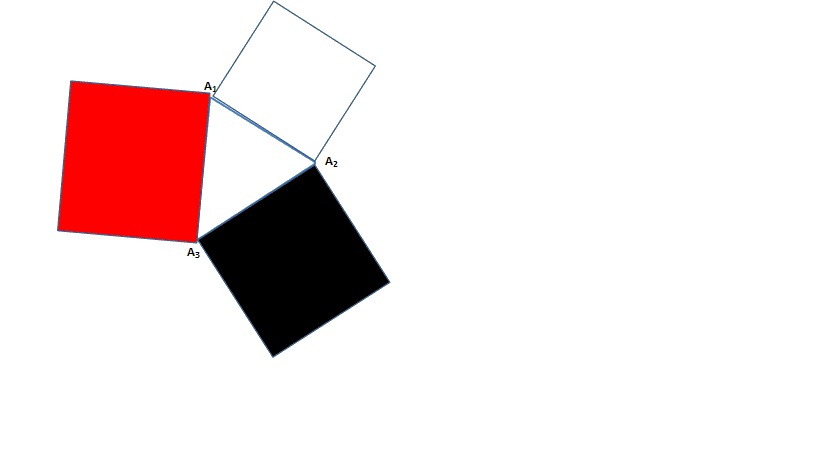}\\
 \caption{Triada of Malevich's squares}
\end{figure}
The sum of the areas of these three squares is expressed in terms of
the three probabilities $p_k$ as follows:
\begin{equation}\label{eq.M4}
S=y_1^2+y_2^2+y_3^2=2\left[3\left(1-p_1-p_2-p_3\right)
+2p_1^2+2p_2^2+2p_3^2+p_1p_2+p_2p_3+p_3p_1\right].
\end{equation}
\noindent The three squares constructed, using the sides of the triangle, are
analogs of the {\it Triada of Malevich's squares}~\cite{Malevich}.
The properties of area $S$ given by Eq.(\ref{eq.M4}) associated with
the triada are different for the classical system states and for the
quantum system states, namely, for three classical coins and for
qubit states.

For classical coins, the numbers $p_1$, $p_2$, and $p_3$ take any
values in the domains $0\leq p_k\leq1$; this means that for
statistics of classical coins the area of the {\it Triada of
Malevich's squares} satisfies the inequality
\begin{equation}\label{eq.M5}
0\leq S\leq6.
\end{equation}
the {\it Triada of Malevich's squares} contains the black square,
the red square, and the white square~\cite{Malevich}. For qubit
states, the probabilities $0\leq p_k\leq1$ to have spin projections
$m=+1/2$ along three orthogonal directions satisfy the uncertainty
relation inequality~(\ref{intq}).

In view of this inequality, the area of the three Malevich's squares
can be equal to neither six nor zero but satisfies the inequality
$S_{\rm min}\leq S\leq S_{\rm max}.$
For all the pure states, $|\psi_z\rangle=\left(\begin{array}{cc}
1\\
0
\end{array}\right),~|\psi_x\rangle=
\frac{1}{\sqrt2}\left(\begin{array}{cc}
1\\
1
\end{array}\right),~\mbox{and}~|\psi_y\rangle=
\frac{1}{\sqrt2}\left(\begin{array}{cc}
1\\
i
\end{array}\right)$
with the parameters ${\bf p}_1=(1/2,1/2,1)$, ${\bf
p}_2=(1,1/2,1/2)$, and ${\bf p}_3=(1/2,1,1/2)$, with the area being
$S=5/2$. If one chooses the probabilities $p_1$, $p_2$, and $p_3$
corresponding to the pure state determined by the point on the Bloch
sphere, which is maximally close to the nearest vertex of the cube,
the value $S=3$ will be obtained. For maximally mixed qubit state
with $p_k=1/2$, the area $S=3/2$. The maximum and minimum values of
the three Malevich's squares correspond to the maximum and minimum
values of the triangle area
\begin{equation}\label{eq.M7}
S_{\mbox{tr}}=\frac{1}{4}\left[\left(y_1+y_2+y_3\right)\left(y_1+y_2-y_3\right)
\left(y_2+y_3-y_1\right)\left(y_3+y_1-y_2\right)\right]^{1/2}.
\end{equation}
The area in the case of three classical coins statistics can take
zero value as the minimum, and the maximum area of the triangle is
equal to ${\sqrt3}/{2}$. For qubit states, the triangle area has
different bounds. The area for the maximally mixed qubit state is
equal to $\sqrt 3/4$; in this case, all three squares of Malevich's
triada are equal and have sides of length unity.

\section{Positive and Completely Positive Maps of the Probabilities
$
{p_1}$, $
{p_2}$, and $
{p_3}$}
In this section, we consider the map of a matrix $\rho$,
\begin{equation}\label{A}
\rho=\left(\begin{array}{cc}
p_3&p^\ast-\gamma^\ast\\
p-\gamma & 1-p_3
\end{array}\right),
\end{equation}
where $p_3^\ast=p_3$, $p=p_1+ip_2$, and $\gamma=({1}/{2})(1+i)$, of
the form
\begin{equation}\label{B}
\rho\longrightarrow\rho_v=\sum_k V_k\rho V_k^\dagger.
\end{equation}
Here, the matrix $V_k$ has the matrix elements
$V_k=\left(\begin{array}{cc}
V_{11}(k)&V_{12}(k)\\
V_{21}(k)&V_{22}(k)
\end{array}
\right)$.
In the case of equalities $\sum_k V_k^\dagger V_k=1$ and
$\rho=\rho^\dagger$, $\mbox{Tr}\rho=1$, and $\rho\geq 0$, the map
given by (\ref{B}) is completely positive map of the density matrix
$\rho$, which provides the new density matrix $\rho_V$, where
$\rho_V=\rho_V^\dagger$, $\mbox{Tr}\,\rho_V=1$, and $\rho_V\geq 0$.
The map of the density matrix can be expressed in terms of the
linear transform of the probabilities $p_1$, $p_2$, and $p_3$, i.e.,
for the vector\\[-11mm]
\begin{equation}\label{D}
{\bf P}=\left(\begin{array}{ccc} p_3\\p\\p^\ast
\end{array}\right),
\end{equation}
the map (\ref{B}) provides the vector ${\bf P}_V$ corresponding to
the density matrix $\rho_V$ of the form
\begin{equation}\label{E}
{\bf P}_V=M_V{\bf P}+{\bf \Delta}_V.
\end{equation}
Here, the 3$\times$3 matrix $M_V$ reads
\begin{equation}\label{F}
M_V=\sum_k\left(\begin{array}{ccc}
|V_{11}(k)|^2-|V_{12}(k)|^2&V_{11}^\ast(k)V_{12}(k)&V^\ast_{12}(k)V_{11}(k)\\
V_{11}^\ast(k)V_{21}(k)-V_{12}^\ast(k)V_{22}(k)&V_{11}^\ast(k)V_{22}(k)&V_{12}^\ast(k)V_{21}(k)\\
V_{21}^\ast(k)V_{11}(k)-V_{22}^\ast(k)V_{12}(k)&V_{21}^\ast(k)V_{12}(k)&V_{22}^\ast(k)V_{11}(k)
\end{array}\right),
\end{equation}
and the vector
\begin{equation}\label{G}
{\bf \Delta}_V=\left(\begin{array}{ccc}
\delta_3\\ \delta\\ \delta^\ast \end{array}\right)
\end{equation}
has three components $\delta_3$, $\delta$, and $\delta^\ast$ of the
form
\begin{eqnarray*}
&&\delta_3=\sum_k\left[|V_{12}(k)|^2-\gamma V_{11}^{\ast}(k)V_{12}(k)-\gamma^\ast
V_{12}^{\ast}(k)V_{11}(k)\right],\\
&&\delta=\sum_k\left[V_{12}^{\ast}(k)V_{22}(k) -\gamma
V_{11}^{\ast}(k)V_{22}(k)-\gamma^\ast
V_{12}^{\ast}(k)V_{21}(k)+\gamma\right].
\end{eqnarray*}
If the map (\ref{B}) is given as the unitary map,
$\rho\longrightarrow\rho_u=u\rho u^\dagger,\quad u u^\dagger=1$,
the transformed probabilities determining the density matrix $\rho_u$ are given by the vector
\begin{equation}\label{S}
{\bf P}_u=M_u{\bf P}+{\bf\Delta}_u,
\end{equation}
where
\begin{equation}\label{Z}
M_u=\left(\begin{array}{ccc}
|u_{11}|^2-|u_{12}|^2&u_{11}^{\ast}u_{12}&u_{12}^{\ast}u_{11}\\
u_{11}^{\ast}u_{21}-u_{12}^\ast u_{22}&u_{11}^{\ast}u_{22}&u_{12}^{\ast}u_{21}\\
u_{21}^{\ast}u_{11}-u_{22}^\ast
u_{12}&u_{21}^{\ast}u_{12}&u_{22}^{\ast}u_{11}
\end{array}\right).
\end{equation}
The shift three-vector ${\bf \Delta}_u$ for the unitary map has the
components $\delta_3(u)$, $\delta(u)$, and $\delta^\ast(u)$ of the
form
\begin{eqnarray}\label{Y}
\delta_3=|u_{12}|^2-\gamma u_{11}^{\ast}u_{12}-\gamma^\ast
u_{12}^{\ast}u_{11},\qquad \delta=u_{12}^{\ast}u_{22} -\gamma
u_{11}^{\ast}u_{22}-\gamma^\ast u_{12}^{\ast}u_{21}+\gamma.
\end{eqnarray}
The 4$\times$4 matrix ${\cal M}_u=\left(\begin{array}{cc}M_u&{\bf \Delta}_u\\
0&1\end{array}\right)$ provides an example of the representation of
the group of unitary 2$\times$2 matrices satisfying the condition
${\cal M}_{u_1}{\cal M}_{u_2}={\cal M}_{u_1u_2}.$

In the process of the qubit evolution determined by the Hamiltonian
$H(t)$, which can be either time-dependent or time-independent, the
density matrix $\rho$ evolves by means of the unitary matrix $u(t)$.
In this case, the evolution of the qubit state can be expressed as a
linear transform of the probabilities $p_3(t)$, $p_1(t)$, and
$p_2(t)$ given by (\ref{S}) with matrix $M_{u(t)}$ and vector ${\bf
\Delta}_{u(t)}$. The matrix elements of the matrix given by
(\ref{Z}) and vector components of the vector given by (\ref{Y})
depend on the matrix elements of the unitary matrix $u_{jk}(t)$,
$j,k=1,2$, if one takes into account the dependence on time.

For stationary Hamiltonian $H$ the evolution is determined by the
unitary matrix
$u(t)=\exp(-i t H).$
If the Hamiltonian $H=H^\dagger$ is
\begin{equation}\label{eq.A}
H=\left(\begin{array}{cc}
H_{11}&H_{12}\\H_{21}&H_{22}\end{array}\right),
\end{equation}
the matrix $u(t)$ reads
\begin{equation}\label{eq.b}
u(t)=\left[\cosh\alpha\left(\begin{array}{cc} 1&0\\0&1\\\end{array}\right)+
\sinh\alpha\left(\begin{array}{cc} n_3&n_1-in_2\\n_1+in_2&-n_3\\
\end{array}\right)\right]\exp\left[-\frac{it}{2}\left( H_{11}+H_{22}\right)\right].
\end{equation}
Here, $\alpha=-i t h $,
\[
h=\left[\left(\frac{H_{11}-H_{22}}{2}\right)^2+\left(\frac{H_{12}+H_{21}}{2}\right)^2
+\left(\frac{iH_{12}-iH_{21}}{2}\right)^2\right]^{1/2}
\]
and
\begin{equation}\label{eq.g}
n_1=\frac{H_{12}+H_{21}}{2h},\quad n_2=\frac{i(H_{12}-H_{21})}{2h},\quad n_3=\frac{H_{11}-H_{22}}{2}.
\end{equation}
For the positive map, one uses the transposition transform
$\rho\longrightarrow\rho^{\mbox{tr}},$ which provides the transform
of the probability
\begin{equation}
{\bf P}\longrightarrow{\bf P}_{\mbox{tr}}=\left(\begin{array}{ccc}
1&0&0\\
0&0&1\\
0&1&0
\end{array}\right)
\left(\begin{array}{ccc}
p_3\\p\\p^\ast\end{array}\right)=\left(\begin{array}{ccc}
p_3\\p^\ast\\p
\end{array}\right).
\end{equation}
The generic positive map gives a convex sum of vectors obtained from vector ${\bf P}$ as
\[
{\bf P}\longrightarrow(cos^2\mu){\bf P}_V+(\sin^2\mu){\bf P}_{V'\mbox{tr}}={\bf P}_{\mbox{pos}}.
\]
Here, $\mu$ is an arbitrary real parameter, ${\bf P}_V $ is given by
(\ref{E}), and ${\bf P}_{V'\mbox{tr}}$ is
\[
{\bf P}_{V'\mbox{tr}}=M_{V'}{\bf P}_{\mbox{tr}}+{\bf \Delta}_{V'},
\]
where the matrix $M_{V'}$ and vector ${\bf \Delta}_{V'}$ are described  by (\ref{F})
and (\ref{G}) determined by the matrices $V_k'$.

For the transposition transform, one has $~p_3\longrightarrow
p_3,\quad p_1\longrightarrow p_1, \quad p_2\longrightarrow1-p_2.$
This transform corresponds to the mirror reflection of the triangle
$A_1A_2A_3$ with respect to the mediana. The transpose matrix
provides the transform of Malevich's squares. The unitary matrix $u$
has the angles $\psi$, $\theta$, and $\phi$ as the parameters, i.e.,
\begin{eqnarray*}
u_{11}=\cos\,({\theta}/{2})\,\exp\big[{i(\phi+\psi)}/{2}\big],
\qquad
u_{12}=\sin\,({\theta}/{2})\,\exp\big[{i(\phi-\psi)}/{2}\big],\\
u_{21}=-\sin\,({\theta}/{2})\,\exp\big[{-i(\phi-\psi)}/{2}\big],
\qquad u_{22}=\cos\,({\theta}/{2})\,\exp\big[{-i(\phi+\psi)}/{2}\big].
\end{eqnarray*}
It is easy to see that the probabilities $p_1,\,p_2,\,p_3$ depend
only on two angles $\theta$ and $\psi$. The evolution of the spin
state means the time dependence of angles or changes in sizes of the
Malevich's squares. Thus, an arbitrary evolution of the qubit state
determined by the map $p_1,\,p_2,\,p_3\longrightarrow
p_1(t),\,p_2(t),\,p_3(t)$ given by Eq.~(\ref{E}) corresponds to the
time dependence of the sum of areas of Malevich's
squares~(\ref{eq.M4}) or the triangle area~(\ref{eq.M7}).\\[-5.1mm]

\section{Conclusions}   
To conclude, we summarize the main results of our study.

We introduced the new geometrical interpretation of qubit states. In
addition to the known picture of the qubit state given by a point in
the Bloch sphere, we showed that the state can be identified with a
triangle or three squares called {\it Triada of Malevich's squares}.
The area of the squares is related to quantum correlations, and the
sum of areas of the squares has bounds. Also the area of the
triangle has a bound. The areas are expressed in terms of three
probabilities of positive spin projections on three orthogonal
directions. The quantum spin-1/2 state and classical state of the
three coins have different characteristics of {\it Triada of
Malevich's squares}. For classical coins, the sum of the areas of
the squares takes the values $0\leq S \leq 6$. For quantum spin
state, the area $S$ is larger than zero and smaller than six.

The positive maps of the qubit states (or gates used in quantum
technologies) are described as linear transforms of three
nonnegative probabilities $p_1$, $p_2$, and $p_3$. Thus, we
introduced a new interpretation of qubit gates as the set of
3$\times$3 matrices and three-vectors, which form a semigroup. The
semigroup of the gates transforms the {\it Triada of Malevich's
squares} into another {\it Triada of Malevich's squares}. Also the
semigroup elements (gates) transform the triangle associated with
the qubit state into a triangle defining another qubit state. Any
time evolution of the qubit state is described by the dynamical
semigroup moving the triangle vertices $A_1$, $A_2$, and $A_3$. We
showed that the transpose transform of the density matrix is
described as the mirror reflection determined by the median of an
equilateral triangle.

\section*{Acknowledgments}      
The formulation of the problem of gates for qubits and the results
of Sec.~6 are due to V. I. Man'ko, who is supported by the Russian
Science Foundation under Project No.~16-11-00084; his work was
performed at the Moscow Institute of Physics and Technology.

\end{document}